\documentclass[runningheads]{llncs}
\usepackage{graphicx}
\usepackage{multirow}
\usepackage{pgfplots}
\usepackage{tikz}
\usepackage{mathrsfs}
\usepackage{amsfonts}
\usepackage[ruled]{algorithm2e}
\usepackage{floatrow}
\newfloatcommand{capbtabbox}{table}[][\FBwidth]

\SetCommentSty{mycommfont}
\renewcommand{\thefootnote}{}

\usepackage{amssymb, amsmath, bm, latexsym,comment}
\usepackage{multirow}
\usepackage{xcolor}
\usepackage{subfigure}
\usepackage{indentfirst}
\usepackage{setspace}
\usepackage{verbatim}
\usepackage{array}
\usepackage{arydshln}
\usepackage[misc]{ifsym} 
\usepackage[colorlinks,linkcolor=blue]{hyperref}
\usepackage{xcolor}
\usepackage[normalem]{ulem} 
\newcommand{\etal}{\textit{et al.}}
\newcommand{\ie}{i.e., }

\begin{document}
	%
\title{Multi-Depth Boundary-Aware Left Atrial Scar Segmentation Network}
	\titlerunning{MDBAnet for LA scar segmentation}
	
		\author{Mengjun Wu \inst{}\thanks{The two authors have equal contributions to the paper.}  \and
		Wangbin Ding $^{\star}$ \and
		Mingjin Yang ${^{(\textrm{\Letter})}}$ \and
		Liqin Huang\inst{}}
	
	
	\authorrunning{Mengjun Wu et al.}
	\institute{College of Physics and Information Engineering, Fuzhou University, Fuzhou, China 
	\email{yangmj5@fzu.edu.cn}
	}
	
	%
	%
	%
	%
	\maketitle              
	\begin{abstract}
Automatic segmentation of left atrial (LA) scars from late gadolinium enhanced CMR images is a crucial step for atrial fibrillation (AF) recurrence analysis. However, delineating LA scars is tedious and error-prone due to the variation of scar shapes. In this work, we propose a boundary-aware LA scar segmentation network, which is composed of two branches to segment LA and LA scars, respectively. We explore the inherent spatial relationship between LA and LA scars. By introducing a Sobel fusion module between the two segmentation branches, the spatial information of LA boundaries can be propagated from the LA branch to the scar branch. Thus, LA scar segmentation can be performed condition on the LA boundaries regions. In our experiments, 40 labeled images were used to train the proposed network, and the remaining 20 labeled images were used for evaluation. The network achieved an average Dice score of 0.608 for LA scar segmentation.
    		
	\end{abstract}
	\renewcommand{\thefootnote}{\arabic{footnote}}
	\setcounter{footnote}{0}
	\section{Introduction}
	Atrial fibrillation (AF) is the most common arrhythmia, occurring at any age, from children to the elderly \cite{heeringa2006prevalence}. Clinically,  catheter ablation (CA) \cite{kirchhof2017catheter} is a widely used invasive procedure for AF treatment, but with a 45\% recurrence rate \cite{balk2010predictors}. 
	Recent studies demonstrated the relationship between the recurrence of AF and left atrial (LA) scars after CA \cite{zghaib2018new,jefairi2019relationship}.
	Late gadolinium enhanced (LGE) cardiac MR  has emerged as one of the promising techniques for imaging LA scars \cite{yang2020simultaneous}. Delineating scarring regions from LGE images could analyze the formation of LA scars, and benefit the monitoring and management of AF patients.
	
	Conventional  scar segmentation methods are mainly based on thresholding, region-growing and  graph-cut algorithms \cite{karim2013evaluation}. 
	Deep-learning (DL) based methods have recently been widely studied for LA scar segmentation tasks. Most DL-based methods explore employing LA or LA walls to improve the scar segmentation. For instance, Chen \etal \cite{chen2018multiview} presented a multi-task segmentation methods, where LA and scars were jointly predicted with an attention model;
	Li \etal \cite{li2020joint,li2022atrialjsqnet} formulated the spatial relationship between LA walls and scars as loss function, which could force the network to focus on objective regions during inference. 
	
	
	Generally, the size of LA scars is varied largely. In the training dataset of LAscarQS 2022 \cite{li2022atrialjsqnet,li2022medical,li2021atrialgeneral},  each LGE image contains average 41.17 scars, and the size of scars are ranged from 0.98 mm$^3$ to 7545.89 mm$^3$. Table \ref{tab:scar_static} presents the statistical information of the scars in LAscarQS 2022. One can observe, 76.1\% of scars sizes are within 50 mm$^3$, and they occupy 16.17\% of total scars volume; whereas only 2.8\% of scars sizes are larger than 500 mm$^3$, but they cover 48\% of total scars volume in the whole dataset. For the tiny objects, a shallower network could outperform the deep U-Net; For the large objects, a deeper network could outperform the shallower network \cite{zhou2019unet++}. The optimal depth of a segmentation network can vary due to the variety of sizes,  which poses an additional challenge in performing scar segmentation.
	
	\begin{table}[]
	    \centering
        \resizebox{1\textwidth}{!}{
	    \begin{tabular}{c|ccccccccccc}
	    \hline
	       Range (mm$^3$) & 0-50 &  50-100 & 100-150 & 150-200 & 200-250  & 250-300 & 300-350 & 350-400 & 400-450& 450-500 & $>$500 \\
	       \hline
	       Number of Scar & 1881&262&121&39&24&17&20&14& 14&7&71\\
	       Percentage (\%) &76.15&10.61&4.89&1.58&0.97&0.69&0.809&0.566&0.566&0.28&2.87\\
	       Total Number & \multicolumn{11}{c}{2470}\\
	       \hline \hline
	       Scar Volume & 33014&18624&14981&6778&5237&4632&6510&5273&5898&3321&99918\\
	       Percentage (\%) &16.17&9.12&7.34& 3.32&2.56&2.27&3.19&2.58&2.89&1.626&48.93\\
	       Total Scar Volume & \multicolumn{11}{c}{204191}\\
	        \hline
	    \end{tabular}
	    }
	    \caption{statistical information of scarring regions in the training dataset of LAScarQS2022}
	    \label{tab:scar_static}
	\end{table}

\begin{figure}[htp]
	    \centering
	    \includegraphics[width=1\textwidth]{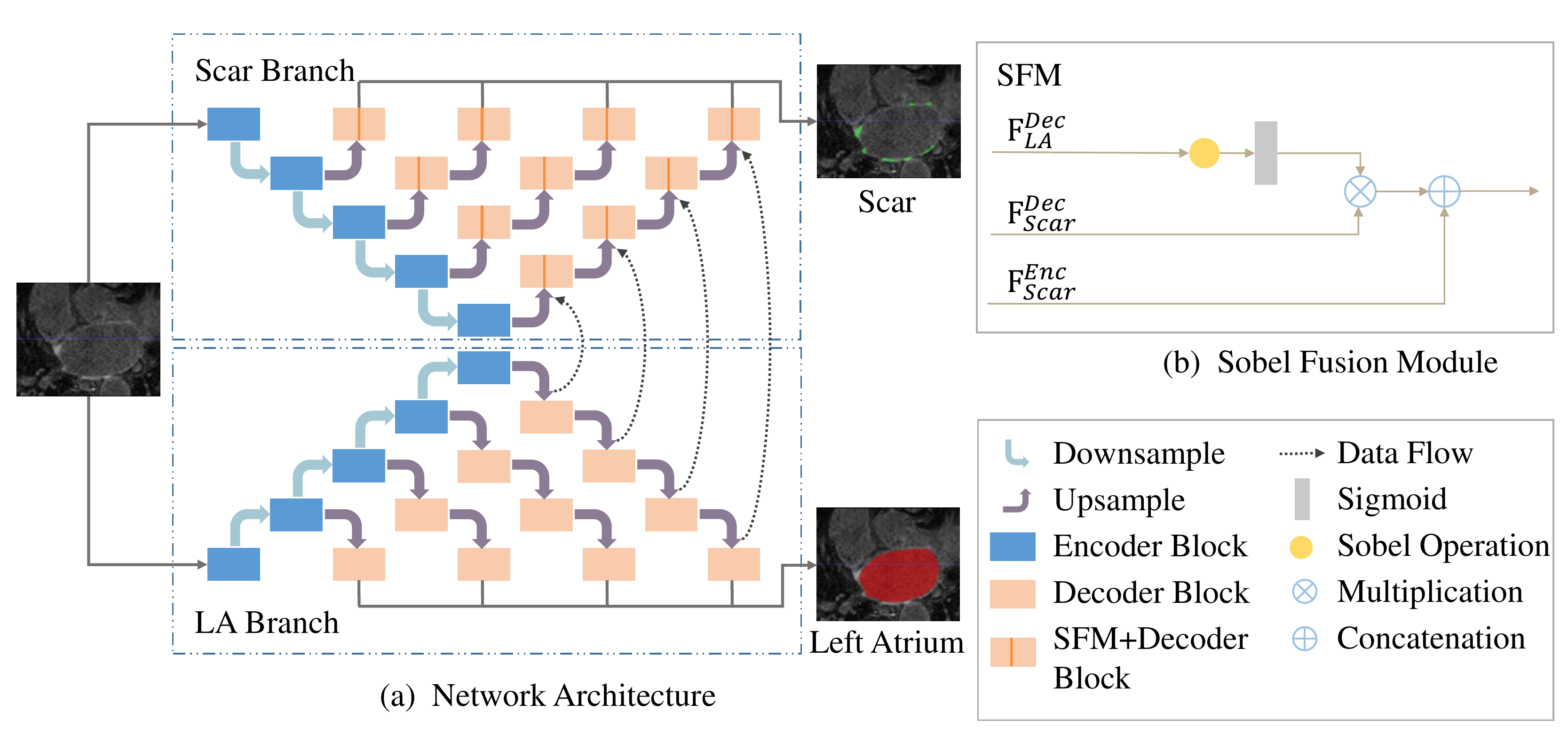}
	    \caption{The architecture of multi-depth boundary-aware network (MDBAnet). It consists of two segmentation branches, \ie the scar branch and the left atrial (LA) branch. In both branches, we introduce multiple U-Nets with different depths to perform scar and LA segmentation.  Furthermore, we propose a Sobel fusion module to extract and propagate LA boundaries information from the LA branch to the scar branch. For conciseness, we only reserve the data flow between the deepest LA decoder path and scar decoder, and all skip connections are omitted.
	    }
	    \label{fig:network}
	\end{figure}

    As shown in Figure \ref{fig:network}, we propose a multi-depth boundary-aware network, namely MDBAnet, to segment different sizes of  LA scars. 
    The main contribution of this work includes:
    (1) We present a multi-depth segmentation network to segment multiple sizes of scars. 
    (2) We propose a plug-and-play Sobel \cite{xu2018cfun} fusion module, which aims to extract LA boundary information to  improve scar segmentation. 

	\section{Method}
	    \subsection{Network Architecture}
	    MDBAnet comprises two branches, which segment LA and LA scars, respectively. 
	    Scars are distributed on the LA wall, and the size of scars varies largely, as seen from our statistical information (Table \ref{tab:scar_static}).
	    In order to achieve scar segmentation of different sizes, the scar branch is stacked with multiple U-Nets that share the same encoder but with different decoder depths. We expect that the shallow networks will focus on segmenting small-size scars, while the deep networks will focus on segmenting large-size scars. Finally, we fuse  segmentation results of each U-Net  as follows:

	    \begin{equation}
	    \hat{Y}_{Scar}=\frac{1}{N} \sum_{n=1}^{N} \hat{Y}_{n}
	    \end{equation}
        where $N$ is the number of  U-Nets with different depths, and $\hat{Y}_{n}$ is the output of the corresponding U-Net.
        
        Furthermore, we aim to improve the performance of scar segmentation by jointly performing LA segmentation. Thus, as shown in Figure \ref{fig:network}, we introduce the LA branch for LA segmentation. The network architecture of the LA branch is symmetric to the scar branch, which is stacked with multiple U-Nets. It outputs LA regions for LGE images. 

       \subsection{Sobel Fusion Module}
\begin{figure}
	    \centering
	    \includegraphics[width=\textwidth]{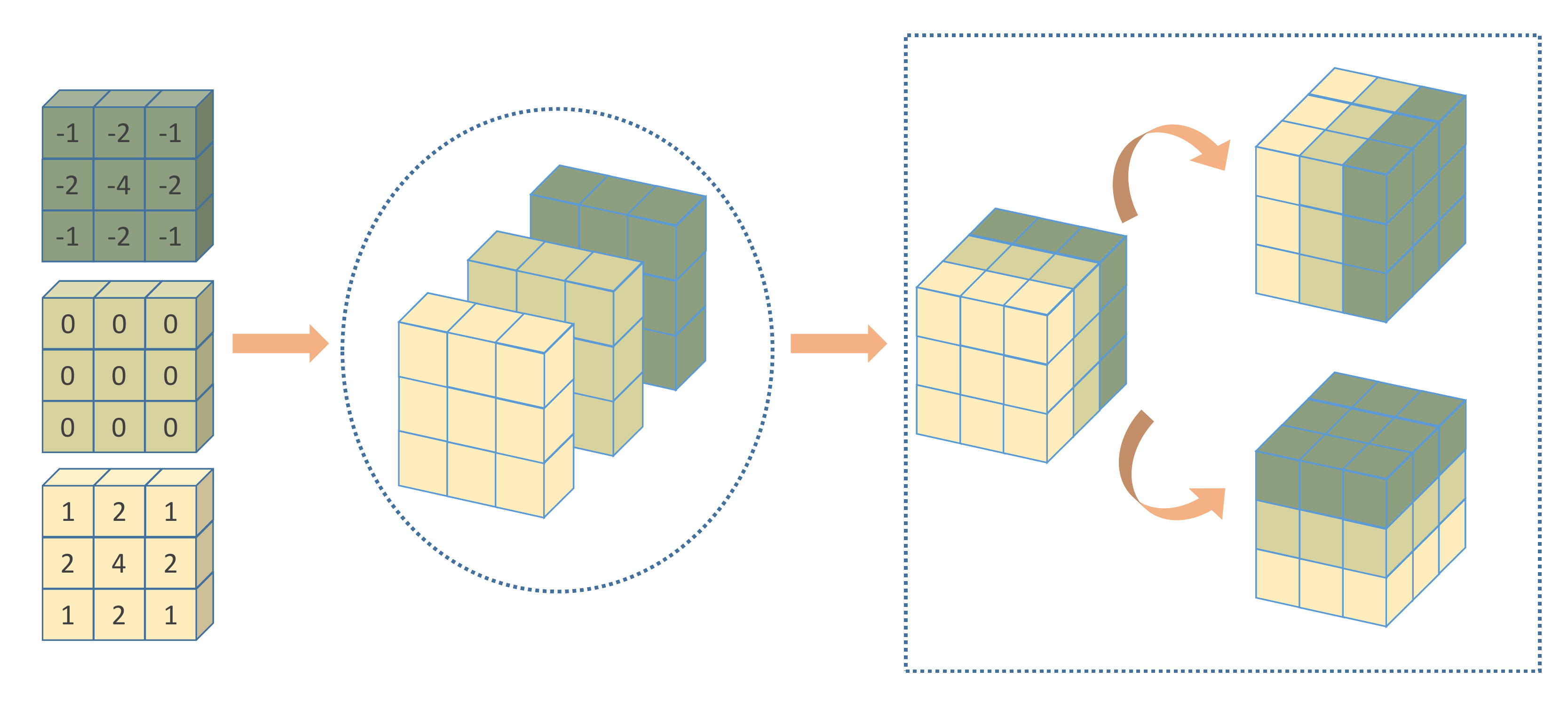}
	    \caption{Our 3D Sobel kernel}
	    \label{kernal}
    	\end{figure}

       We explore the inherent spatial relationship between LA and LA scars. Generally, LA scars are  distributed around  LA boundaries. We introduce a Sobel \cite{kanopoulos1988design} operator to extract the boundary information of feature maps.  Then a Sobel fusion module (SFM) is proposed  to take full advantage of the spatial relationship between boundary information and LA scars. The input of SFM includes the feature maps of LA decoder, the previous layer of scar decoder and scar encoder. The output of SFM  can be calculated as follows:
	 \begin{equation}
	 {F}_{}^{out}=\left(F^{Dec}_{Scar} \otimes S\left(F^{Dec}_{LA}\right)\right) \oplus F^{Enc}_{Scar},
	 \end{equation}
	where $\otimes$ and $\oplus$ represent element-wise multiplication and concatenation, respectively, $S$ represents 3D Sobel operation, $F^{Enc}$ and $F^{Dec}$ are the feature map from the encoder path and decoder path, respectively. Here, Sobel operation is implemented by a fixed kernel convolution layer, which consists of three 3D Sobel kernels. Following Xu \etal \cite{xu2018cfun}, each 3D Sobel Kernel can be described as a 3×3×3 matrix, as shown in  Figure \ref{kernal}. They can be used to extract the boundary information from the axial, sagittal and coronal views of image. 

	Based on the SFM,  the feature map from the decoder path of the LA branch could be passed to the 3D Sobel Kernel to get the boundary information of LA. Then we re-calibrate the feature map of the scar branch with the boundary information, which provides spatial attention and forces the network to focus on the LA boundaries region. 
	
	\subsection{Loss Function}
    We employ Dice loss and cross-entropy loss  to jointly optimize the segmentation results of the network.
	The total loss of our network is:
	 
	 \begin{equation}
	 \begin{split}
	 \mathcal{L}=&-DCS(\hat{Y}_{Scar},Y_{Scar})+CE(\hat{Y}_{Scar},Y_{Scar})\\
	 &-DCS(\hat{Y}_{LA},Y_{LA})+CE(\hat{Y}_{LA},Y_{LA}),
	 \end{split}
	 \end{equation}
	where $\hat{Y}_{\{Scar,LA\}}$ and $Y_{\{Scar,LA\}}$ are the predicted and gold standard labels, respectively; $DCS(a,b)$ calculate the Dice score (DS) between a and b; and $CE(a,b)$ calculate the cross-entropy loss.
	
	 
	

	\section{Experiment}
      \subsection{Dataset}
   We trained and evaluated our method on the Left Atrial and Scar Quantification \text { \& } Segmentation Challenge 2022 (LAScarQS 2022) dataset, which aimed to segment LA and LA scars from LGE CMR images. The challenge dataset provides a total of  60 labeled and 10 unlabeled LGE CMR images, and gold standard labels include: LA  and LA scars. 
   In our experiment, we split the labeled images into a training set of 40 cases, and the remaining 20 cases for evaluation. Finally, the performance of the network was evaluated on 10 unlabeled images.
   
    \subsection{Implementations}
    Our network was implemented in PyTorch, using two NVIDIA GeForce RTX 3080 GPUs. We used SGD optimizer to adjust the network parameters (batch size=2, weight decay=0.00003, momentum=0.99). The initial learning rate was set 0.01 and decayed exponentially. During training, enhancement techniques, \ie random rotation, random scaling, random elastic deformation, gamma-corrected enhancement and mirroring, were applied on the fly. 
    

   \subsection{Result}
   We compared our method to three different segmentation methods: 
   \begin{itemize}
    \item nnU-Net \cite{isensee2018nnu}: One of the state-of-the-art segmentation networks. We trained it with 3D LGE images as well as corresponding scar labels. 
    \item MDnet: A multi-depth segmentation network based on U-Net, which is the scar branch of MDBAnet.
    \item MDBAnet$_{mul}$: A variation of  MDBAnet.  We implement a multiplication fusion module to propagate information from the LA branch to the scar branch.
    \item MDBAnet: The proposed network. It consists two branches with multi-depth network to segment LA and scars. We implement a Sobel fusion module to propagate information from the LA branch to the scar branch.
    \end{itemize}
   To evaluate methods, DS and Hausdorff distance (HD)  were calculated between the prediction results and the gold standard label.
   
	\begin{table}[]
	    \centering
        \resizebox{1\textwidth}{!}{
	    \begin{tabular}{l|cc|cc}
	    \hline
	        \multirow{2}{4em}{Methods}  & \multicolumn{2}{c|}{LA}& \multicolumn{2}{c}{Scar}  \\
	        \cline{2-5}
	          & DS & HD (mm) &DS& HD (mm) \\
	       \hline
	       nnU-Net & -&-&0.488(0.090)&39.62(12.81)\\
	       MDnet & -&-&0.504(0.085)&40.66(12.88)\\
	      MDBAnet$_{mul}$ &\textbf{0.926(0.020)}&\textbf{17.83(11.33)}&0.504(0.087)&33.21(10.43)\\
	       MDBAnet&0.923(0.027)&19.18(11.10)&\textbf{0.512(0.083)}&\textbf{31.67(10.86)}\\
	        \hline
	    \end{tabular}}
	    
	    \caption{The performance of different methods. DS: Dice score; HD: Hausdorff distance. Note that MDBAnet and MDBAnet$_{mul}$ jointly produce scar and LA segmentation, we thus list the LA segmentation results of this two methods.}
	    \label{tab:scar_result}
	\end{table}
	
	Table \ref{tab:scar_result} shows the segmentation performance of different methods.
    Methods using a multi-depth strategy (\ie MDBAnet, MDBAnet$_{mul}$ and MDnet ) had obtained better segmentation performance compared to single-depth network, \ie nnU-Net. Particularly, MDnet improved Dice and HD by 0.01 and 1 for scar segmentation, respectively.
    It proves the  effectiveness of using multi-depth strategy.
    Meanwhile, scar segmentation could be further improved by jointly performing LA segmentation. One can see that both MDBAnet$_{mul}$ and MDBAnet improve the scar segmentation results by utilizing information from the LA branch. For example,  MDBAnet achieved an improvement DS by 0.8\% (p$=$0.125), and significantly reduced HD from 40.66 to 31.67 (p$<$0.05) against MDnet. 
    Besides, MDBAnet$_{mul}$ propagated the entire feature maps of the LA branch, while the MDBAnet extracted the boundary information via SFM. MDBAnet could obtain better DS and HD for scar segmentation. This implied the benefit of SFM.
    
\begin{figure}
	    \centering
	    \includegraphics[width=\textwidth]{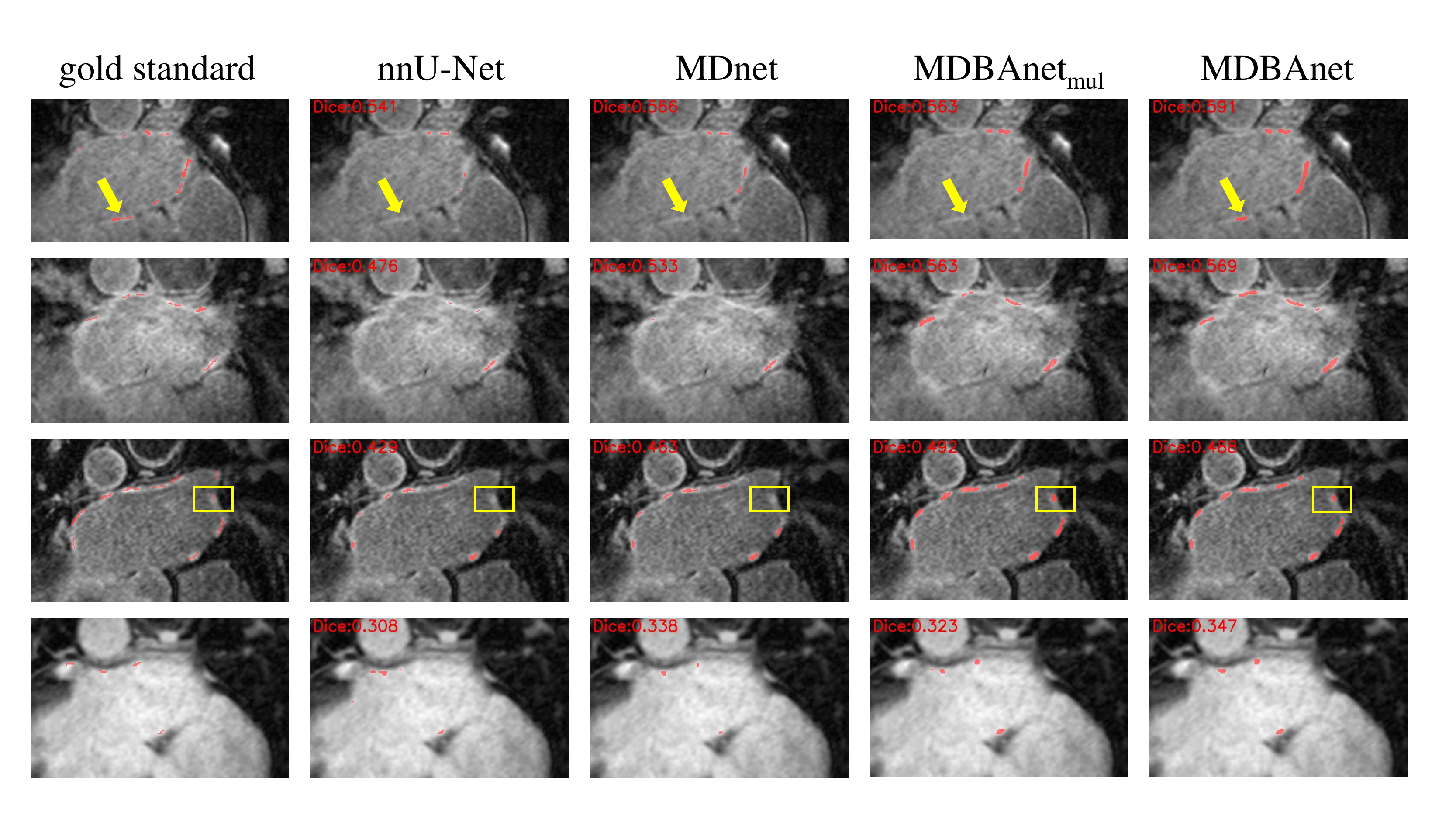}
	    \caption{Visualization  of different scar segmentation methods. Yellow Arrows mark the advantage of MDBAnet, while yellow Boxes denote the tiny scars.}
	    \label{scar}
    	\end{figure}

    In Figure \ref{scar}, we showed four typical cases for visualization. nnU-Net may failed to perform segmentation for tiny scars (yellow Boxes), which is consist to the quantanity result of Table \ref{tab:scar_result}. Moreover,  MDBAnet achieved better results for some difficult cases, such as ambiguity scars (yellow Arrows). This was probably due to the usage of SFM, which could force the scar branch to focus on boundary regions.

	\section{Conclusion}
	In this work, we  have proposed a multi-depth boundary-aware LA scar segmentation network. It consists of two segmentation branches based on multi-depth strategy. Meanwhile, we implemented a SFM to propagate information from  LA branch to  scar branch. The experimental results showed that multi-depth network has a positive effect on scar segmentation, and  SFM was capable of further improving scar segmentation performance. The network achieved a DS of 0.608 on validation data of LAScarQS 2022.

\bibliographystyle{splncs04}

\bibliography{ref}

\begin{thebibliography}{10}
\providecommand{\url}[1]{\texttt{#1}}
\providecommand{\urlprefix}{URL }
\providecommand{\doi}[1]{https://doi.org/#1}

\bibitem{balk2010predictors}
Balk, E.M., Garlitski, A.C., ALSHEIKH-ALI, A.A., Terasawa, T., Chung, M., Ip,
  S.: Predictors of atrial fibrillation recurrence after radiofrequency
  catheter ablation: a systematic review. Journal of cardiovascular
  electrophysiology  \textbf{21}(11),  1208--1216 (2010)

\bibitem{chen2018multiview}
Chen, J., Yang, G., Gao, Z., Ni, H., Angelini, E., Mohiaddin, R., Wong, T.,
  Zhang, Y., Du, X., Zhang, H., et~al.: Multiview two-task recursive attention
  model for left atrium and atrial scars segmentation. In: International
  Conference on Medical Image Computing and Computer-Assisted Intervention. pp.
  455--463. Springer (2018)

\bibitem{heeringa2006prevalence}
Heeringa, J., van~der Kuip, D.A., Hofman, A., Kors, J.A., van Herpen, G.,
  Stricker, B.H.C., Stijnen, T., Lip, G.Y., Witteman, J.C.: Prevalence,
  incidence and lifetime risk of atrial fibrillation: the rotterdam study.
  European heart journal  \textbf{27}(8),  949--953 (2006)

\bibitem{isensee2018nnu}
Isensee, F., Petersen, J., Klein, A., Zimmerer, D., Jaeger, P.F., Kohl, S.,
  Wasserthal, J., Koehler, G., Norajitra, T., Wirkert, S., et~al.: nnu-net:
  Self-adapting framework for u-net-based medical image segmentation. arXiv
  preprint arXiv:1809.10486  (2018)

\bibitem{jefairi2019relationship}
Jefairi, N.A., Camaioni, C., Sridi, S., Cheniti, G., Takigawa, M., Nivet, H.,
  Denis, A., Derval, N., Merle, M., Laurent, F., et~al.: Relationship between
  atrial scar on cardiac magnetic resonance and pulmonary vein reconnection
  after catheter ablation for paroxysmal atrial fibrillation. Journal of
  Cardiovascular Electrophysiology  \textbf{30}(5),  727--740 (2019)

\bibitem{kanopoulos1988design}
Kanopoulos, N., Vasanthavada, N., Baker, R.L.: Design of an image edge
  detection filter using the sobel operator. IEEE Journal of solid-state
  circuits  \textbf{23}(2),  358--367 (1988)

\bibitem{karim2013evaluation}
Karim, R., Housden, R.J., Balasubramaniam, M., Chen, Z., Perry, D., Uddin, A.,
  Al-Beyatti, Y., Palkhi, E., Acheampong, P., Obom, S., et~al.: Evaluation of
  current algorithms for segmentation of scar tissue from late gadolinium
  enhancement cardiovascular magnetic resonance of the left atrium: an
  open-access grand challenge. Journal of Cardiovascular Magnetic Resonance
  \textbf{15}(1),  1--17 (2013)

\bibitem{kirchhof2017catheter}
Kirchhof, P., Calkins, H.: Catheter ablation in patients with persistent atrial
  fibrillation. European heart journal  \textbf{38}(1),  20--26 (2017)

\bibitem{li2020joint}
Li, L., Weng, X., Schnabel, J.A., Zhuang, X.: Joint left atrial segmentation
  and scar quantification based on a dnn with spatial encoding and shape
  attention. In: International Conference on Medical Image Computing and
  Computer-Assisted Intervention. pp. 118--127. Springer (2020)

\bibitem{li2021atrialgeneral}
Li, L., Zimmer, V.A., Schnabel, J.A., Zhuang, X.: Atrialgeneral: Domain
  generalization for left atrial segmentation of multi-center lge mris. In:
  International Conference on Medical Image Computing and Computer-Assisted
  Intervention. pp. 557--566. Springer (2021)

\bibitem{li2022atrialjsqnet}
Li, L., Zimmer, V.A., Schnabel, J.A., Zhuang, X.: Atrialjsqnet: A new framework
  for joint segmentation and quantification of left atrium and scars
  incorporating spatial and shape information. Medical Image Analysis
  \textbf{76},  102303 (2022)

\bibitem{li2022medical}
Li, L., Zimmer, V.A., Schnabel, J.A., Zhuang, X.: Medical image analysis on
  left atrial lge mri for atrial fibrillation studies: A review. Medical Image
  Analysis p. 102360 (2022)

\bibitem{xu2018cfun}
Xu, Z., Wu, Z., Feng, J.: Cfun: Combining faster r-cnn and u-net network for
  efficient whole heart segmentation. arXiv preprint arXiv:1812.04914  (2018)

\bibitem{yang2020simultaneous}
Yang, G., Chen, J., Gao, Z., Li, S., Ni, H., Angelini, E., Wong, T., Mohiaddin,
  R., Nyktari, E., Wage, R., et~al.: Simultaneous left atrium anatomy and scar
  segmentations via deep learning in multiview information with attention.
  Future Generation Computer Systems  \textbf{107},  215--228 (2020)

\bibitem{zghaib2018new}
Zghaib, T., Nazarian, S.: New insights into the use of cardiac magnetic
  resonance imaging to guide decision making in atrial fibrillation management.
  Canadian Journal of Cardiology  \textbf{34}(11),  1461--1470 (2018)

\bibitem{zhou2019unet++}
Zhou, Z., Siddiquee, M.M.R., Tajbakhsh, N., Liang, J.: Unet++: Redesigning skip
  connections to exploit multiscale features in image segmentation. IEEE
  transactions on medical imaging  \textbf{39}(6),  1856--1867 (2019)

\end{thebibliography}

\end{document}